# Evidence of the production of hot hydrogen atoms in RF plasmas by catalytic reactions between hydrogen and oxygen species

Jonathan Phillips[*] and Chun Ku Chen,

Department of Chemistry and Nuclear Engineering, University of New Mexico,

Albuquerque, NM

and Randell Mills,

BlackLight Power, Cranbury, NJ

Abstract

Selective H-atom line broadening was found to be present throughout the volume (13.5 cm ID x 38 cm length) of RF generated $H_2O$ plasmas in a GEC cell.  Notably, at low pressures (ca. <0.08 Torr), a significant fraction (ca. 20%) of the atomic hydrogen was "hot" with energies greater than 40 eV with a pressure dependence, but only a weak power dependence.  The degree of broadening was virtually independent of the position studied within the GEC cell, similar to the recent finding for $He/H_2$ and $Ar/H_2$ plasmas in the same GEC cell.  In contrast to the atomic hydrogen lines, no broadening was observed in oxygen species lines at low pressures.  Also, in "control": $Xe/H_2$ plasmas run in the same cell at similar pressures and adsorbed power, no significant broadening of atomic hydrogen, Xe, or any other lines was observed.  Stark broadening or acceleration of charged species due to high electric fields can not explain the results since (i) the electron density was insufficient by orders of magnitude, (ii) the RF field was essentially confined to the cathode fall region in contrast to the broadening that was independent of position, and (iii) only the atomic hydrogen lines were broadened. Rather, all of the data is consistent with a model that claims specific, predicted, species can act catalytically through a resonant energy transfer mechanism to create "hot" hydrogen atoms in plasmas.

---

[*] To whom correspondence should be addressed

**INTRODUCTION**

Plasma sources have been developed over decades as light sources, ionization sources for mass spectroscopy, excitation sources for optical spectroscopy, and sources of ions for surface etching and chemistry. Perhaps surprisingly, only in the last decade has extensive spectroscopic characterization been conducted on "mixed gas" plasmas, and these studies revealed surprising observations for those mixed gas plasmas in which hydrogen was one of the gases. Specifically, in mixtures of argon and hydrogen, the hydrogen emission lines are significantly broader than any argon or molecular hydrogen line. The broadening is universally believed to show that many of the hydrogen atoms in the plasma are at super elevated temperatures (>200,000 K), and are in fact much hotter than any other species, including electrons (ca. 5,000K) present in the plasma. These readily repeatable observations primarily for $Ar/H_2$ plasmas [1-15], but for other mixed gas plasmas as well [16-21], now play a role in a debate regarding the very fundamentals of physics.

On one side of a developing controversy are those who postulate the energy for the selective heating is derived from field acceleration (FA) of hydrogen ions. On the other side is a new theory of quantum physics, Classical Quantum Mechanics (CQM), based entirely on Maxwell's and Newton's Laws. This model postulates that the energy for selective H atom heating comes from a catalytic chemical process in which the electron of hydrogen, postulated to be a spherical shell of charge (physical object of precise size and shape), 'shrinks' to a new diameter determined by solutions to a Newtonian force balance. This process is predicted only to take place in specified plasma gas mixtures. Perhaps surprisingly the data collected to date appears to be far more consistent with the latter theory than the former. The present work was designed to further test the two models. And it is argued in the *Discussion* that the results of the current data set, as well as all others, are consistent with the predictions of the CQM hypothesis and not with those of the FA models.

Kuraica and Konjevic [1, 2], Videnovic et al. [3] are among the first to postulate FA models. Specifically, they postulated that the energy required to create the anomalously "hot" (>20 eV) hydrogen in RF plasmas studied by their team was generated by the acceleration of hydrogen ions such as $H^+$, $H_2^+$, and $H_3^+$ in the high fields (e. g. over 10 kV/ cm) present in the cathode fall region. Subsequent charge exchange and dissociation of the field-accelerated ions left "hot" neutral H atoms. Djurovic and Roberts [5] recorded the spectral and spatial profiles of Balmer α line emission from low pressure RF (13.56 MHz) discharges in $H_2$ plus Ar mixtures in a direction normal to the electric field. The introduction of Ar in a pure $H_2$ plasma increased the number of fast neutral atoms as evidenced by the intensity of the broad component of a two-component Doppler-broadened Balmer α line profile.

Extensive and detailed work in the same GEC cell employed in the present study with Ar/$H_2$, as well as He/$H_2$ plasmas [15,16] is qualitatively consistent with their observation that independent of cell position or direction of observation relative to the applied field, the average energy of a wide profile component was 23.8 eV for voltages above 100 V, and the average energy of a slow component was 0.22 eV. The mechanism proposed by Djurovic and Roberts is the production of fast H atoms from electric field accelerated $H_2^+$. The explanation of the role of Ar in the production of a large number of excited hydrogen atoms in the $n= 3$ state, as well as raising their energy for a given pressure and applied RF voltage, is that collisions with Ar in the plasma sheath region enhances the production of fast $H_2$ from accelerated $H_2^+$. The fast $H_2$ then undergoes dissociation to form fast H which may then be excited locally to the $n= 3$ state by a further collision with Ar. The local excitation is a requirement since the atomic lifetime of the hydrogen $n=3$ state is approximately $10^{-8}$ s, and the average velocity of the hydrogen atoms is < $10^5$ m/$s$. Thus, the distance traveled must be less that 0.001 m.

The models described above, do not explain the observation that the shapes of the broadened Balmer series lines are independent of the perspective of observation relative to the applied field, as well as evidence that the $n=3$ state is not preferential populated. This prompted the development of a number of modified FA models. The modifications require hot H to form when high energy $H^+$, $H_2^+$ or $H_3^+$ ions [6] bombard an electrode surface rich in absorbed hydrogen, leading to the subsequent ejection of hot hydrogen atoms [7]. Yet, not only does this mechanism still appear incapable of explaining the completely symmetric shape of the broadened lines, independent of the direction of observation relative to the applied field, there is no evidence presented, nor presumably found, for the existence of 'hot' ionic species.

More recently, there has been some attempt to explain the repeated observation of line broadening at positions far from the field in terms of 'indirect' absorption of energy from the field. Specifically, it was postulated that electrons in the high end of the EEDF directly transfer energy to the translational energy mode of atomic hydrogen [14] in low field regions of the discharge. This is clearly implausible as electron temperature is known to strongly couple to the atomic excitation temperature, and very weakly to translation temperature [22]. Thus, the measurement of excitation temperature in the present study and in others of about 0.5 eV, is not consistent with electrons generating a translational motion temperature (of only one species!) almost two orders of magnitude greater than the excitation temperature.

Balmer series line broadening has also been repeatedly reported for mixed gas plasmas containing hydrogen, but not argon. Preferential hydrogen line broadening, often extreme, was found in a number of mixed gas discharge plasmas, particularly He/$H_2$ [16–21]. In none of these plasmas was there any significant broadening of the noble gas lines. Furthermore, hydrogen lines were not broadened in a number of other mixed gas plasmas including Xe/$H_2$ (10%), and Kr/$H_2$ (10%). These results show that the presence of hot hydrogen in mixed gas plasmas is not limited to Ar/$H_2$. The results are consistent with predictions of CQM [8–13, 17,23,24], that only very special catalytic species will react in plasmas with hydrogen by a resonant, nonradiative energy transfer mechanism to generate hot hydrogen. Species such as $He^+$, $Ar^+$, and $Sr^+$ meet the catalyst criterion–a chemical or physical process with an enthalpy change equal to an integer multiple of $E_h$= 27.2 eV where $E_h$ is one hartree. Catalyst are identifiable on the basis of their known electron energy levels. Conversely, species such as atoms or ions of Xe do not fulfill the criterion.

The present paper is part of a series of efforts by our team to study selective line broadening in GEC/RF plasmas for the purpose of determining if the FA or CQM models better predict the outcome. That is, in the best tradition of science, we have designed experiments for which the predictions of two different paradigms are dramatically different. Specifically, this is the third in a series of studies of 'resonant transfer' (RT) mixed gas plasmas predicted by CQM to produce line broadening throughout the plasma glow, and not just in the high field region. In contrast, the FA models predict broadening only in high field regions, if at all. The two earlier studies, one of Ar/$H_2$ and one of He/$H_2$ plasmas produced results completely consistent with the predictions of CQM, (and hence inconsistent with FA model predictions) including nearly constant magnitude broadening throughout the plasma and independent of direction relative to that of the applied field, insensitivity to applied field strength, neutral H-atom temperatures more than an order of magnitude hotter than that of electrons, and no Balmer broadening in non-RT mixtures. The present results are also far more consistent with the outcome predicted by the CQM model.

**EXPERIMENTAL**

**Plasma Hardware**

All plasmas were generated in a GEC-type cell [6,7,25] held at 0.5 Torr. This system, shown elsewhere [16], consisted of a large cylindrical (14 cm ID x 36 cm length) Pyrex chamber containing two parallel steel circular (8.25 cm diameter) plates, placed about 1 cm apart at the center. RF power from a RF VII, Model RF 5 13.6 MHz power supply was sent to the plates through 8 mm diameter steel feeds, which entered the chamber through standard Ultratorr fittings, one on each end of the chamber. UHP grade (99.999%) $H_2$ and *Xe* gases were metered into the chamber through Ultratorr fittings at one end, about 18 cm from the electrodes, using two mass flow controllers (MKS).

Water vapor was generated by pumping on a reservoir (approx. 20 cc) of distilled, de-ionized water. The flow rate was not directly controlled, but rather a needle valve was adjusted to maintain the desired pressure, as measured by a MKS Baratron placed above a Welch two-stage rotary vane oil-sealed vacuum pump (Model 8920) with a rated capacity of 218 l/min. This pump was attached to the chamber with a 1 cm ID Ultratorr fitting at the end opposite that at which gas entered. All parts, chamber, power supply, gauges, spectrometer, etc. were grounded with heavy-duty Reynolds aluminum foil to improve the magnitude of the signal to noise ratio.

**Spectrometer**

The spectrometer system used in this study, described in detail elsewhere [26], was built around a 1.25 m visible light instrument from Jobin Yvon-Spex with a holographic ruled diffraction grating (1800 g/mm), with a nearly flat response between 300 and 900 nm, and the slit was set at 10 μ*m* in all cases. Light was collected using a light fiber bundle consisting of 19 fibers, each of 200 μ*m* diameter, and a CCD for a detector. Light was input to the spectrometer from the light fiber placed at position (1) near the inlet end of the chamber approximately 15 cm from the cathode, (2) in a quartz insert tube 1 cm in diameter that ended about 1 cm from the edge of the electrodes, or (3) near the pump end of the chamber approximately 15 cm from the anode.

The fiber was oriented "orthogonal" relative to the azimuthal axis of the chamber in all cases. It is important to note that tests with a red laser with the system open clearly showed that light emanating from the region between the parallel-plate electrodes could not have reached the "hooded" fiber optic probe when it was positioned at either end. Indeed, our tests showed that light emitted at least 14 cm from the plate region reached the fiber optic probe at Positions 1 and 3. These readings were consistent with the listed 9° acceptance angle of the probe corresponding to a 1 cm diameter "spot." Moreover, the probes were oriented such that the acceptance cone should "miss" the power feeds by several centimeters.

In most cases, the data used for computations (e.g. excitation temperatures) was collected for the same time over the same wavelength region. Balmer series spectral lines were fit using three Gaussian curves: one for the "cold" (<0.15 eV) hydrogen, one for "warm" (<2 eV) hydrogen, and the third for "hot" (>10 eV) hydrogen. It is notable that the fittings achieved were excellent, producing $R^2$>0.98 in all cases. One reason for the excellent fits was the absence of any signal in the relevant spectral region of the Balmer α and β lines of the water plasma (Figure 1a). In contrast, there was "signal" in the same regions for the $Xe/H_2$ (10%) plasma (Figure 1c).

## RESULTS

A significant amount of data was collected in order to reliably detect trends in the H-atom line broadening in water plasmas as a function of plasma operating conditions. Data on line broadening was systematically collected for the Balmer $\alpha$, $\beta$, $\gamma$, and $\delta$ lines at the three positions given in the Experimental section. Measured values of H-atom $\alpha$ and $\beta$ line broadening for water plasmas at Positions 2 and 3, at ten or more pressures, and at three absorbed power levels (100 W, 150 W and 200 W), are presented in Tables I–VI. The data for the $\gamma$ and $\delta$ lines is not presented as it is considered less reliable as the intensity of these lines is significantly lower. Also, the trends in Doppler energy of the hot hydrogen as a function of pressure and applied power is virtually identical to those observed from the $\alpha$ and $\beta$ lines; thus, the data for the higher energy transitions in the Balmer series is regarded as somewhat redundant, and hence not essential to the arguments presented.

Typical peaks and the best fit of the data to three Gaussian curves corresponding to Doppler broadening are shown in Figure 1. All mechanisms other than Doppler would not produce a three component line, rather only a single component line.

The pressure and position dependence for a given adsorbed power plotted in Figures 2 and 3 shows that the energy of the hot hydrogen (as well as cold and warm atomic hydrogen) is independent of position, but strongly dependent on the pressure, dropping sharply above approximately 0.1 Torr in all cases. A comparison between the plots also suggests only a weak dependence on the absorbed energy. In sum, the results reported here clearly show that there is "hot" atomic hydrogen, of an apparent energy between 40 and 55 eV, independent of position, from the $H_\alpha$ line (approx. 450,000 to 550,000 K) and nearly 70 eV from the $H_\beta$ line throughout

GEC cell water vapor plasmas generated at low pressures (ca. < 0.08 Torr). No other species in these plasmas, specifically molecular hydrogen and various oxygen species, were found to be "hot" at these low pressures.

Under all operating conditions the magnitude of the broadening at 15 cm from the electrodes was very close to the magnitude within the region between the plates. It is notable that the fraction of H-atoms that are hot is somewhat impacted by position within the cell. As shown in Figure 4, the fraction of "hot" hydrogen is strongly dependent on pressure, and somewhat on adsorbed power, and is generally slightly higher between the plates than it is at the end of the cell, 15 cm from the electrodes.

Simple comparisons can be made between the results of this study of $H_2O$ plasmas and earlier studies, performed in the same GEC cell, of $He/H_2$ plasmas [16]. First, it is notable that there is no overlap in the Doppler energies of the hot hydrogen measured for the two different plasmas, despite the fact that the physical arrangement of the cell (e.g. electrode separation) was virtually identical in both cases. For the water plasma the average broadening is always at least 10 eV greater than that found in the $He/H_2$ plasma. Also, one feature similar to that found with the $He/H_2$ plasma, is that the hydrogen concentration is asymmetric. For example, for the water plasmas it was consistently about twice as high at the "pump end" (Position 3) as it is at the inlet end (Position 1). Although an exhaustive study of hydrogen lines was not made at Position 1, due to the low intensity of H-atom emission at that position, a limited number of comparisons were made, and it was clear that the magnitude of the line broadening, the average excitation, and other features were very nearly identical at both ends of the cell. Another difference: most (ca. 80%) of the atomic hydrogen was "hot" in the $He/H_2$ plasma; whereas, less than half of the atomic hydrogen was found to be hot in the water plasma.

The intensities of all four Balmer lines were used to obtain a measure of the average hydrogen excitation temperature at Positions 2 and 3 as given in Figure 5 and Table VII. The excitation temperature was found to be around $0.5 \pm 0.1$ eV (approx. $5000 \pm 1000$ K) for all of the plasmas. This temperature was independent of the source of the different Balmer lines used in the computation within the error range: relative intensities of the cold, hot, or total hydrogen components of the Balmer lines. These values are also similar to electron temperatures measured in earlier studies of low pressure Ar plasmas generated at slightly lower powers in a large glass cavity [27]. It must be noted that it is clear that in plasmas of the sort studied here the excitation temperature and the electron temperature are closely coupled. Thus, the electron temperature throughout the cell is clearly of the order 0.5 eV (*22*).

To test the catalyst mechanism, a control plasma, $Xe/H_2$, was studied in some detail in this same GEC cell. Just as in the earlier studies [15,16], the control plasmas produced only narrow Balmer series lines (<2.5 eV) away from the electrodes and some "warm" hydrogen (<3 eV) between the electrodes as shown in Figure 1C. In the earlier studies, the "control" plasma was run at pressures similar to those of the $He/H_2$ plasma studied, about 0.5 Torr. For this study, the control plasmas were studied at pressures close to those at which selective hydrogen broadening in a water plasma was observed (<0.9 Torr). The Balmer series line intensities in $Xe/H_2$ plasmas were very low even though the data collection times were four times greater than those for the water plasmas at matched pressures. Thus, it was necessary to run nearly pure $H_2$ to get sufficient signal at the low pressures employed in the present work.

**DISCUSSION**

In mixed gas plasmas containing argon and hydrogen, selective line broadening of atomic hydrogen lines (no broadening of lines belonging to argon, molecular hydrogen, helium or ions of any type) in high field regions [1-21] has been reported repeatedly. Even in pure hydrogen plasmas, generated with DC discharge or RF systems [14], including one group using a GEC cell [7], selective broadening of atomic hydrogen lines has been reported.

All groups agree that the broadening of the lines is Doppler in origin. Stark broadening can be eliminated because the required electron densities are orders of magnitude greater than the gas densities. Moreover, the lines are composed of three parts: hot, warm and cold. All H atoms, not just a fraction, as well as other species, would be impacted by high charge densities. Optical thickness cannot be a factor by the same argument: the entire line would be broadened, not just a fraction. Computation also shows that the optical thickness cannot be a factor. Specifically, for optically thin plasmas (self adsorption not significant), the effective path length $\tau_\omega(L)$ is less than one:

$$\tau_\omega(L) = \sigma_\omega N_H L < 1 \qquad (1)$$

where $\sigma_\omega$ is the absorption cross section, $N_H$ is the number density, and L is the plasma path length traversed by the light. The absorption cross section for Balmer α emission is $\sigma = 1 \times 10^{-16}$ cm [28]. An upper limit on the excited $H_\alpha$ density, assuming all of the water is fully dissociated, the temperature (as measured) is 5000 K, and, the excited states are populated according to the Boltzmann distribution (as measured), is $10^3$ cm$^{-3}$. No more than 15 cm of plasma is traversed. Putting these values together yields an effective path length of the order $10^{-12}$ cm. Clearly, these plasmas are optically thin. Other potential explanations such as instrument broadening can be readily eliminated because those mechanisms would not produce selective broadening of one species. Moreover, all the Balmer series lines are broadened approximately to the same energy level, a result completely consistent with Doppler broadening.

All of the above arguments apply to the line broadening observed in the present work. Thus, we conclude that some of the hydrogen atoms (between 10 and 45 percent, Figure 4) in the water plasma are selectively "heated" to extremely high temperatures, 450,000 to 700,000 K. About half of the remaining hydrogen ("cold") produces line broadening consistent not with a Doppler effect, but rather with a combination of Stark effect, instrument effects, etc. A third type of hydrogen ("warm") may be due to a catalytic effect of hydrogen alone when maintained in high concentration such found on the surface of the cathode as reported previously [9, 13].

**Inconsistency with FA Models**

As discussed in the Introduction the standard physics models for the generation of the hot hydrogen in $Ar/H_2$ plasmas are all field acceleration models (FA) as they all require that the hydrogen ions obtain energy directly from the field [1-7, 14, 29-31]. The possibility that the hot hydrogen forms from collisions with hot gas species (ions or atoms) is considered highly unlikely. Such models would require a thermal equilibrium between all species, and this is not observed. Spectroscopy indicates that no hot argon of any form is present [15, 30]. Thus, the selective H heating is explained only in terms of acceleration of charged H species.

There are two classes of FA models: (1) FA models postulating a gas phase mechanism for formation of hot hydrogen near the electrodes, and (2) FA models requiring that an H ionic species hits the electrode resulting in energy transfer to absorbed hydrogen species and consequently the desorption of a hot hydrogen atom. In the "bombardment" models hydrogen species on electrode surfaces are "hit" by energized ions, generally $H_3^+$, $H_2^+$, or $H^+$ ions [3], and subsequently ejected as hot hydrogen [30, 31] or these species recoil and disintegrate to fast H.

In all cases, these models only predict selective hydrogen broadening in high field regions. In the gas production model, a hydrogen ion such as $H_3^+$ [3] that is increased in concentration by interactions of $H_2$ with Ar, is accelerated by the field toward an electrode, captures an electron via interaction with an Ar, dissociates to form *n*=3 state hydrogen, or forms n=3 state hydrogen via collision with a neutral Ar [30], and then emits.

There does not appear to be any variation on those FA models capable of explaining the observations of the present work. First, hot hydrogen was found throughout the chamber rather than only in the vicinity of the electrodes. Hot hydrogen ions created near the electrodes simply cannot migrate 15 cm without equilibrating with the plasma gas. Given a maximum computed mean free path of the hot hydrogen of 0.5 cm at 0.1 Torr, the high temperature would have to remain undiminished through hundreds of collisions to be observed at 15 cm distance from the electrode. The excitation temperature of the "parent" atomic hydrogen species was only about 5000 K (approximately 0.5 eV), and the excitation temperature of RF plasmas is generally associated with the electron temperature [22, 32, 33]. This means that the internal temperature of the atomic hydrogen, as well as the temperature of the electrons in the plasma, were about two orders of magnitude lower than that of the hot hydrogen. Thus, any FA model must explain how H atoms can be two orders of magnitude hotter than the electrons in the GEC plasmas studied.

Even relatively obscure postulated processes were considered as mechanisms to provide the observed energy of the hot hydrogen atoms. For example, the Frank-Condon effect [34-37] could create "hot" neutrals with energies between 2 and 4.5 eV via wall reactions of the type:

$$3H \rightarrow H_2 + H(4.5 \text{ eV}) \qquad (2)$$

where the third body (wall) removes the bond energy.

The postulated energies are always less than 5 eV, hence, the energy of neutral species created in this fashion do not match the energies of the neutrals observed in this study.

In sum, it is untenable to suggest modifications of the earlier FA models can explain the present data. For example, all earlier models require acceleration of ions in the high field (unscreened) regions near the electrodes. The earlier models also include other specific predictions, such as preferential population of $n=3$ states, which are not observed. The gas phase models must be rejected for two additional reasons. First, they all require a high cross section for charge transfer, peculiar to argon and hydrogen ions, to allow for the rapid charge transfer necessary to create neutral, high energy $H_2$, which must be formed before high energy (neutral) H atoms can form. There was no argon in the plasmas studied for this work. Second, the $H_2$ lines were not observed to be broadened. Third, it is not plausible to suggest that the fields found 15 cm from the electrode are as strong as those found in the boundary layer near the electrodes. Field screening by the sheath reduces the fields dramatically within millimeters, and a highly conductive plasma bulk is essentially a very low equipotential [6]. Yet, the hot hydrogen found at 15 cm from the electrodes was of the same energy as that found between the electrodes. This third objection to the gas phase models clearly also shows the "bombardment" models to be implausible. Indeed, how can a hot hydrogen atom generated at the electrode by bombardment traverse 15 cm of the plasma without loosing energy or thermalizing? Clearly, the electrode bombardment models are not consistent with the finding of the present study that the degree of broadening was the same throughout the plasma volume. Even more problematic for this directional "migrating" fast H is the symmetry of the broadened profile which requires non-directionality of the fast H.

Moreover, we cannot identify any previously proposed mechanism that can produce hydrogen atoms with an average energy of about 45 eV for the water plasma, nearly twice the average energy observed for He/$H_2$ plasmas generated in an identical system [16]. Why is this process completely absent in the Xe/$H_2$ plasmas? What "energy from the field" process would produce neutrals with energies two orders of magnitude higher than those of the electrons in the plasma?

**Consistency with CQM**

These results extend the list of plasmas generated in a GEC cell with RF power, that have been thoroughly tested to determine if the selective heating of H atoms, independent of field and applied voltage, predicted by the CQM theory (so-called 'rt-plasmas') takes place. The results were again positive. Specifically, the results of this work are among the first [20, 38] to show that in oxygen containing plasmas, even in regions far from any significant accelerating field, superheated H atoms can be found. Studies of GEC systems are particularly valuable as the systems are widely available, and our experience is that the data obtained in these systems is remarkably consistent. Thus, GEC data should be easy to replicate in any lab.

Our team has already participated in studying GEC plasmas and producing data of relevance to the debate for Ar/$H_2$ [15] and He/$H_2$ [16] systems. Our most recent study shows that in in low pressure (0.5 Torr) He/$H_2$ (10%) RF plasmas maintained in a GEC cell only the spectral lines of the H-atoms are Doppler broadened. That is, there was no broadening at all of the He lines, or molecular hydrogen lines. Moreover, the hydrogen Balmer lines were broadened consistently to the same magnitude ~27 eV throughout the volume of the cell, and not simply in the region between the electrodes. The finding that the broadening was found throughout the volume indicated that earlier explanations for selective H-atom broadening in Ar/H$_2$ RF plasmas were not applicable. The earlier explanations all require that the origins of the Doppler energy is acceleration of H ions in the vicinity of the electrodes, even in cases were the excess broadening was found to be independent of position in the inter-electrode region [5, 6]. Another finding in that study, also consistent with the rt-plasma model, was that in Xe/H$_2$ (10%) plasmas there was no broadening of either hydrogen or Xe lines outside of the electrode region. Also, even though the voltage between the electrodes was changed, by more than a factor of four, no concomitant change in the magnitude of the broadening was observed anywhere in the plasma. This is clearly inconsistent with any reasonable expectation of a field acceleration model. The conclusion of that study was that so many aspects of the data were consistent with a "chemical" reaction occurring between He$^+$ and hydrogen species throughout the volume of the cell, and so many aspects completely inconsistent with any of the various FA models, that the data was strong evidence of the superiority of the CQM model.

It is also to be pointed out that there have been prior reports of line broadening and related phenomenon associated with atomic hydrogen spectral lines, both Balmer and Lyman (requiring EUV spectroscopy), in oxygen containing plasmas [39, 40], although not in a GEC type system. These results clearly cannot be explained with FA models that require Ar to be present.

**Proposed CQM Mechanism**

The CQM postulated catalytic reaction requires atomic hydrogen and atomic or molecular oxygen. These species are favored to form at low pressures, which is consistent with the observed pressure dependence. Under plasma discharge [41-44] and the inherent photolysis conditions [45] water is known to undergo decomposition to primarily hydrogen atoms and hydroxyl radials, and the hydroxyl radicals can further form hydrogen and oxygen atoms.

The average energy of the fast H, according to CQM, depends on the particular catalyst as well as the conditions of the reaction. As observed here, the broadening is not a function of field strength, or field direction. As observed, broadening is expected to be observed throughout the plasma and will be isotropic.

The particular mechanism of the broadening in the water plasma is determined by the bond and ionization energies of the oxygen species. Specifically the bond energy of $O_2$ is 5.165 eV and the first, second, and third ionization energies of an O-oxygen atom are 13.61806 eV, 35.11730 eV, and 54.9355 eV, respectively [46]. Thus, the reactions:

$$O_2 \rightarrow O + O^{2+} \tag{3}$$

$$O_2 \rightarrow O + O^{3+} \tag{4}$$

and

$$2O \rightarrow 2O^{+} \tag{5}$$

provide a net enthalpy of about 2, 4, and 1 times $E_h$, respectively, and hence all of the above processes are candidates for the catalytic generation of hydrinos, according to the CQM theory. For example, the following two body (postulated) catalytic reaction will produce (metastable) hydrinos:

$$H(a_H) + O_2 \rightarrow O + O^{2+} + H^*(a_H/3) + 54.4 \text{ eV} \tag{6}$$

Where H($a_H$) is a 'ground state' hydrogen atom as conventionally understood, and H*($a_H$/3) is a 'metastable' form of a H($a_H$/3) hydrino. H($a_H$/3) is a stable hydrino (shrunken atomic hydrogen) with a diameter 1/3 that of H($a_H$), and H($a_H$) has a radius of one Bohr radius.

This proposed reaction meets the postulated catalytic requirements of CQM. To wit: the process affecting the catalytic agent (i.e. dissociation and ionization of the oxygen molecule) requires $2 \times 27.2$ eV which for this case is approximately half the energy of the transition (54.4 eV, per reaction (4)) to the lower (smaller) state of the hydrogen electron (108.8 eV). Hence, the postulated catalytic processes invariably create metastable hydrino species with a large 'excess' of energy. In particular, the metastable form of hydrino created in Rxn (4) carries 54.4 eV of 'excess' energy. This is quickly lost via one of the following reactions:

$$\text{H*}(a_H/3) \rightarrow \text{H}(a_H/3) + h\nu \quad (7)$$

Or

$$\text{H*}(a_H/3) + \text{H}(a_H) \rightarrow \text{Fast H}(a_H/3) + \text{Fast H}(a_H) \quad (8)$$

The former will create a 54.4 eV photon and the latter a fast hydrogen with a translational energy of 27.7 eV. Why is the fast photon emitted with twice the energy transferred to the atomic hydrogen? The reaction must conserve linear momentum. As written (eq. 8) both species enter the process with virtually no momentum, and thus must leave the reaction point with equal and opposite momentum. As both species have the same mass, they must also leave with equal energy.

It is interesting to note that without exception the data from the GEC shows that ALL of the energy absorbed by the H($a_H$) is absorbed as translational as predicted. That is, the proposed process as written is consistent with the observation that the excitation temperature of the hot H atoms observed is orders of magnitude lower than the translational energy.

Other molecular oxygen 'catalytic' reactions conform to the rules of CQM. Thus, molecular oxygen acting as a catalyst can also produce $H(a_H/2)$ species:

$$H(a_H) + 2O \rightarrow 2O^+ + H^*(a_H/2) \tag{9}$$

Once again, the metastable species quickly decays, either releasing a photon (13.6 eV), or reacting with another species (e.g. $H(a_H)$) in the plasma to produce 'hot' species:

$$H^*(a_H/2) + H(a_H) \rightarrow \text{Fast } H(a_H/2) + \text{Fast } H(a_H) \tag{10}$$

In this case each species leaves the reaction with 6.6 eV of kinetic energy.

Other processes involve the reaction of one hydrino with another:

$$H(a_H/3) + H(a_H/3) \rightarrow H(a_H/2) + H(a_H/4) + 27.2 \text{ eV} \tag{11}$$

Where the 27.2 eV can either be given off as kinetic energy (e.g. Rxn 8) or as a photon (Rxn 7). The above reaction is representative of a set of processes termed 'disproportionation' [21, 23, 47]. These reactions are anticipated in the CQM as the hydrinos themselves have appropriate 'energy holes' to act as catalytic agents. Clearly starting with any set of hydrinos, produced catalytically per reactions such as (6) and (9), it is anticipated that lower energy state hydrinos, down to the lowest state (H(1/137)), can be produced via disproportionation processes. Each step 'down' to a smaller (literally) hydrogen species releases more energy, thus, the amount of energy available for creating high kinetic energy hydrogen atoms gets very large. Thus, the metastable species created via disproportionation reactions are key to forming H atoms with energies in excess of 40 eV.

Another set of processes involving two reacting hydrinos, can create hydrogen atoms. For example:

$$H(a_H/3) + H(a_H/2) \rightarrow H(a_H) + H(a_H/4) + 54.4 \text{ eV} \tag{12}$$

It is possible that the metastable hydrino decomposes quickly enough that the $H(a_H)$ produced in reaction (12) absorbs 27.2 eV of kinetic energy. Moreover, if the plasma contains a high density of 'hot' hydrinos, likely given the metastable decomposition processes posited (e.g. reactions (7), (10)), hydrino/hydrino reactions (e.g. reactions (11) and (12)) will create metastables of very high energies that upon collisional decay can produce atomic hydrogen with translational energies far higher than 27 eV. Such reactions can readily produce H atoms with energies greater than 45 eV, as observed.

Given the proposed mechanisms, why do we not observe hot O atoms, or hot $O_2$? That is, why don't we see evidence of metastable hydrinos decomposing via reaction with species other than atomic hydrogen? CQM indicates that a metastable species will form during the catalytic process, and further predicts that the energy transfer will have a much higher cross section with H since it is a resonant process. Efficient energy transfer can occur by resonant collision mechanisms including dipole–dipole coupling, wherein the angular frequency of the electrons of any hydrogen are related by integers. We can further employ experimental evidence to determine likely routes to metastable decay. On the basis of observation we offer a three part answer to the question of why there is an obvious preference for the production of hot H atoms and not hot oxygen species of any type. First, the cross section for metastable decay to produce hot species via 'collision' with H($a_H$) is much higher than that with any other species. The experimental evidence clearly indicates the reaction, with H atoms is dominant. Possibly the cross section for metastable decay via conversion of energy to translational energy is only significant for this type of H*–H collisions. Collisions of the metalstable with all other species, we suggest, would be unfavorable, and the drop in energy to the true stable state in the absence of atomic hydrogen will be via photon emission. Second, energy-transferring collisions between metastables and heavier species, with a concomitant drop in energy to the stable state, even if they do occur, will not produce a 'visible' effect. In a collision with a 50 eV hydrino, an O atom would absorb at most 3 eV, and $O_2$ only half of that. Moreover, the increase in velocity of species, and hence the relative change in line width goes as the square root of the mass ratio. Thus, relative to hydrogen, the absolute increase in line width of atomic oxygen (assuming the same emission frequency) is only one-fourth. Third, inelastic collisions would lead to the 'overheating' of internal modes, such as electron excitation or molecular dissociation, rather than overheating of just the translational mode. The spectroscopic evidence of this would be far harder to find than the spectacular line broadening that accompanies the energy transfer to the translational mode of atomic hydrogen.

Finally, it should be noted that the phenomenon of direct (no photon) transfer of energy from metastable species to atomic species is not unique to CQM theory. In fact there are several reports in the literature of selective and direct transfer of energy from metastable noble gases (helium or neon) to hydrogen atoms in excited gases [48-50]. In these cases no line broadening is observed, and the 'resonant energy transfer' of energy clearly selectively excites one electronic transition in the hydrogen. For example, near atmospheric pressure admixture of Ne with very low hydrogen concentrations, excited with ionizing particle beams, results in very intense Lyman α radiation, with no concomitant Balmer lines, etc. [48]. This suggests a 10.2 eV resonant energy transfer from excited $Ne_2^*$ to H atoms. As discussed in these reports, the excitations observed are at the far end of what is energetically possible. The observations in this report of hydrogen kinetic energies in excess of 40 eV are far outside the possible range for this postulated mechanism.

**SUMMARY**

Direct mapping of the broadening of Balmer series lines for pure water plasmas in a GEC cell, and the impact of applied power and operating pressure was studied for the first time. It was found that hydrogen atoms were superheated (e.g. >40 eV for pressures less than 0.08 Torr) relative to all other species (e.g. approx. 0.5 eV) everywhere in a large GEC-RF plasma. The findings that super heated hydrogen atoms are found throughout the volume of a GEC cell (not only in high field regions) with an energy independent of position, and direction (i.e. not correlated with field direction) once again indicates that FA models are not capable of explaining the selective heating of H-atoms in plasma systems. In contrast, the broadening of H atoms in a water plasma is consistent with/predicted by a new model of quantum mechanics (CQM). According to this model, a pure water plasma should be an environment favorable to the conversion of H atoms via a catalytic resonant transfer process to a 'shrunken' (hydrino) state. This process will be accompanied by the release of energy that can be preferentially transferred to the kinetic energy mode of H atoms. No heating of H atoms was observed in a $Xe/H_2$ plasma in the same cell, a result also consistent that the rt-process requires specific catalytic species.

Thus, the evidence collected in this study adds to a growing body of experimental data that is consistent with CQM and not consistent with standard quantum physics. Earlier evidence was obtained from line broadening (see Introduction and Discussion), calorimetric [51], extreme ultraviolet (23,52) and nuclear magnetic resonance [53] studies. The fundamentals of the theory of CQM can be found elsewhere [54].

Table 1. Analysis of Hα lines in 100 W water plasmas

| RF Power = 100W | | | | | |
|---|---|---|---|---|---|
| Position / Balmer Lines | Pressure (torr) | Doppler Energy Cold H (eV) | Doppler Energy Warm H (eV) | Doppler Energy Hot H (eV) | Area Ratio Hot/All |
| 2 / H$_\alpha$ | 0.23 | 0.122 | 1.336 | 11.6 | **0.250** |
| 2 / H$_\alpha$ | 0.19 | 0.123 | 1.373 | 10.8 | **0.269** |
| 2 / H$_\alpha$ | 0.17 | 0.122 | 1.412 | 11.0 | **0.288** |
| 2 / H$_\alpha$ | 0.15 | 0.123 | 1.470 | 11.8 | **0.287** |
| 2 / H$_\alpha$ | 0.13 | 0.124 | 1.502 | 12.6 | **0.291** |
| 2 / H$_\alpha$ | 0.12 | 0.122 | 1.459 | 12.5 | **0.303** |
| 2 / H$_\alpha$ | 0.10 | 0.121 | 1.475 | 14.1 | **0.287** |
| 2 / H$_\alpha$ | 0.09 | 0.121 | 1.517 | 16.2 | **0.263** |
| 2 / H$_\alpha$ | 0.08 | 0.119 | 1.560 | 23.1 | **0.245** |
| 2 / H$_\alpha$ | 0.07 | 0.118 | 1.517 | 35.0 | **0.245** |
| 2 / H$_\alpha$ | 0.06 | 0.117 | 1.291 | 37.9 | **0.252** |
| 2 / H$_\alpha$ | 0.05 | 0.116 | 1.193 | 39.2 | **0.251** |
| 2 / H$_\alpha$ | 0.04 | 0.114 | 1.195 | 42.1 | **0.218** |
| 2 / H$_\alpha$ | 0.03 | 0.114 | 1.360 | 42.4 | **0.159** |
| 2 / H$_\alpha$ | 0.02 | 0.104 | 1.539 | 37.3 | **0.078** |
| 3 / H$_\alpha$ | 0.18 | 0.124 | 1.043 | 8.2 | **0.135** |
| 3 / H$_\alpha$ | 0.17 | 0.118 | 1.124 | 12.4 | **0.137** |
| 3 / H$_\alpha$ | 0.15 | 0.122 | 1.136 | 15.1 | **0.127** |
| 3 / H$_\alpha$ | 0.13 | 0.117 | 1.176 | 23.2 | **0.133** |
| 3 / H$_\alpha$ | 0.11 | 0.120 | 1.244 | 20.7 | **0.126** |
| 3 / H$_\alpha$ | 0.09 | 0.118 | 1.218 | 19.1 | **0.141** |
| 3 / H$_\alpha$ | 0.08 | 0.118 | 1.261 | 34.4 | **0.136** |
| 3 / H$_\alpha$ | 0.07 | 0.116 | 1.263 | 23.3 | **0.135** |
| 3 / H$_\alpha$ | 0.06 | 0.118 | 1.371 | 29.7 | **0.156** |
| 3 / H$_\alpha$ | 0.05 | 0.115 | 1.350 | 37.1 | **0.211** |
| 3 / H$_\alpha$ | 0.04 | 0.115 | 1.252 | 37.6 | **0.250** |
| 3 / H$_\alpha$ | 0.03 | 0.113 | 1.432 | 32.2 | **0.083** |

Table 2. Analysis of Hα lines in 150 W water plasmas

| RF Power = 150W | | | | | |
|---|---|---|---|---|---|
| **Position / Balmer Lines** | **Pressure (torr)** | **Doppler Energy Cold H (eV)** | **Doppler Energy Warm H (eV)** | **Doppler Energy Hot H (eV)** | **Area Ratio Hot/All** |
| 2 / $H_\alpha$ | 0.20 | 0.126 | 1.497 | 12.7 | **0.258** |
| 2 / $H_\alpha$ | 0.17 | 0.121 | 1.446 | 12.6 | **0.294** |
| 2 / $H_\alpha$ | 0.15 | 0.123 | 1.536 | 14.0 | **0.300** |
| 2 / $H_\alpha$ | 0.13 | 0.123 | 1.600 | 15.9 | **0.294** |
| 2 / $H_\alpha$ | 0.11 | 0.119 | 1.478 | 17.1 | **0.304** |
| 2 / $H_\alpha$ | 0.09 | 0.118 | 1.562 | 27.1 | **0.281** |
| 2 / $H_\alpha$ | 0.08 | 0.119 | 1.594 | 43.7 | **0.303** |
| 2 / $H_\alpha$ | 0.07 | 0.117 | 1.404 | 44.1 | **0.358** |
| 2 / $H_\alpha$ | 0.06 | 0.117 | 1.110 | 48.9 | **0.345** |
| 2 / $H_\alpha$ | 0.04 | 0.117 | 1.163 | 49.0 | **0.278** |
| 2 / $H_\alpha$ | 0.03 | 0.113 | 1.323 | 52.1 | **0.203** |
| 2 / $H_\alpha$ | 0.02 | 0.111 | 1.687 | 50.7 | **0.145** |
| 3 / $H_\alpha$ | 0.20 | 0.119 | 1.161 | 19.1 | **0.159** |
| 3 / $H_\alpha$ | 0.18 | 0.124 | 1.250 | 25.5 | **0.148** |
| 3 / $H_\alpha$ | 0.15 | 0.126 | 1.167 | 20.9 | **0.160** |
| 3 / $H_\alpha$ | 0.13 | 0.119 | 1.262 | 24.4 | **0.165** |
| 3 / $H_\alpha$ | 0.11 | 0.118 | 1.325 | 23.5 | **0.145** |
| 3 / $H_\alpha$ | 0.09 | 0.115 | 1.410 | 43.7 | **0.189** |
| 3 / $H_\alpha$ | 0.08 | 0.116 | 1.554 | 41.5 | **0.277** |
| 3 / $H_\alpha$ | 0.07 | 0.117 | 1.404 | 44.4 | **0.359** |
| 3 / $H_\alpha$ | 0.05 | 0.115 | 1.378 | 45.2 | **0.335** |
| 3 / $H_\alpha$ | 0.04 | 0.117 | 1.376 | 43.6 | **0.260** |

Table 3. Analysis of Hα lines in 200 W water plasmas

| RF Power = 200W | | | | | |
|---|---|---|---|---|---|
| Position / Balmer Lines | Pressure (torr) | Doppler Energy Cold H (eV) | Doppler Energy Warm H (eV) | Doppler Energy Hot H (eV) | Area Ratio Hot/All |
| 2 / $H_\alpha$ | 0.22 | 0.123 | 1.369 | 11.6 | **0.268** |
| 2 / $H_\alpha$ | 0.18 | 0.125 | 1.459 | 12.4 | **0.296** |
| 2 / $H_\alpha$ | 0.15 | 0.122 | 1.520 | 14.3 | **0.313** |
| 2 / $H_\alpha$ | 0.13 | 0.122 | 1.618 | 17.5 | **0.310** |
| 2 / $H_\alpha$ | 0.12 | 0.121 | 1.668 | 23.7 | **0.306** |
| 2 / $H_\alpha$ | 0.11 | 0.119 | 1.605 | 36.3 | **0.307** |
| 2 / $H_\alpha$ | 0.09 | 0.115 | 1.295 | 51.6 | **0.414** |
| 2 / $H_\alpha$ | 0.07 | 0.117 | 1.174 | 52.1 | **0.432** |
| 2 / $H_\alpha$ | 0.05 | 0.116 | 1.142 | 52.0 | **0.350** |
| 2 / $H_\alpha$ | 0.04 | 0.116 | 1.252 | 53.3 | **0.260** |
| 2 / $H_\alpha$ | 0.03 | 0.110 | 1.585 | 53.2 | **0.159** |
| 2 / $H_\alpha$ | | | | | |
| 3 / $H_\alpha$ | 0.21 | 0.119 | 1.108 | 13.9 | **0.171** |
| 3 / $H_\alpha$ | 0.17 | 0.121 | 1.149 | 14.3 | **0.172** |
| 3 / $H_\alpha$ | 0.13 | 0.123 | 1.158 | 15.0 | **0.181** |
| 3 / $H_\alpha$ | 0.11 | 0.124 | 1.225 | 20.7 | **0.188** |
| 3 / $H_\alpha$ | 0.09 | 0.119 | 1.252 | 23.2 | **0.204** |
| 3 / $H_\alpha$ | 0.08 | 0.120 | 1.370 | 31.2 | **0.203** |
| 3 / $H_\alpha$ | 0.08 | 0.117 | 1.490 | 44.8 | **0.256** |
| 3 / $H_\alpha$ | 0.07 | 0.116 | 1.421 | 50.4 | **0.412** |
| 3 / $H_\alpha$ | 0.05 | 0.116 | 1.352 | 50.6 | **0.421** |
| 3 / $H_\alpha$ | 0.04 | 0.115 | 1.432 | 47.4 | **0.244** |
| 3 / $H_\alpha$ | 0.03 | 0.122 | 2.045 | x | **0.000** |

Table 4. Analysis of Hβ lines in 100 W water plasmas

| RF Power = 100W | | | | | |
|---|---|---|---|---|---|
| Position / Balmer Lines | Pressure (torr) | Doppler Energy Cold H (eV) | Doppler Energy Warm H (eV) | Doppler Energy Hot H (eV) | Area Ratio Hot/All |
| 2 / $H_\beta$ | 0.21 | 0.143 | 1.644 | 11.0 | **0.203** |
| 2 / $H_\beta$ | 0.17 | 0.156 | 1.936 | 25.7 | **0.172** |
| 2 / $H_\beta$ | 0.16 | 0.239 | 2.480 | 14.7 | **0.175** |
| 2 / $H_\beta$ | 0.14 | 0.132 | 1.820 | 19.7 | **0.216** |
| 2 / $H_\beta$ | 0.12 | 0.142 | 1.944 | 18.4 | **0.198** |
| 2 / $H_\beta$ | 0.11 | 0.175 | 3.331 | 17.0 | **0.148** |
| 2 / $H_\beta$ | 0.10 | 0.134 | 1.841 | 13.9 | **0.219** |
| 2 / $H_\beta$ | 0.08 | 0.127 | 1.699 | 22.6 | **0.193** |
| 2 / $H_\beta$ | 0.07 | 0.127 | 1.730 | 28.6 | **0.207** |
| 2 / $H_\beta$ | 0.06 | 0.128 | 1.679 | 38.8 | **0.219** |
| 2 / $H_\beta$ | 0.05 | 0.129 | 1.679 | 39.5 | **0.176** |
| 2 / $H_\beta$ | 0.04 | 0.128 | 1.648 | 46.8 | **0.148** |
| 2 / $H_\beta$ | 0.03 | 0.110 | 1.698 | 22.4 | **0.070** |
| 3 / $H_\beta$ | 0.20 | 0.138 | 1.191 | 5.7 | **0.142** |
| 3 / $H_\beta$ | 0.16 | 0.145 | 1.421 | 8.2 | **0.104** |
| 3 / $H_\beta$ | 0.14 | 0.137 | 1.324 | 5.3 | **0.030** |
| 3 / $H_\beta$ | 0.12 | 0.138 | 1.454 | 26.6 | **0.107** |
| 3 / $H_\beta$ | 0.10 | 0.132 | 1.367 | 12.0 | **0.099** |
| 3 / $H_\beta$ | 0.09 | 0.127 | 1.396 | 6.9 | **0.099** |
| 3 / $H_\beta$ | 0.08 | 0.123 | 1.446 | 14.2 | **0.124** |
| 3 / $H_\beta$ | 0.07 | 0.126 | 1.505 | 11.5 | **0.111** |
| 3 / $H_\beta$ | 0.06 | 0.120 | 1.810 | 52.3 | **0.251** |
| 3 / $H_\beta$ | 0.05 | 0.122 | 1.890 | 58.9 | **0.176** |
| 3 / $H_\beta$ | 0.03 | 0.122 | 1.728 | 63.4 | **0.092** |

Table 5. Analysis of Hβ lines in 150 W water plasmas

| RF Power = 150W | | | | | |
|---|---|---|---|---|---|
| **Position / Balmer Lines** | **Pressure (torr)** | **Doppler Energy Cold H (eV)** | **Doppler Energy Warm H (eV)** | **Doppler Energy Hot H (eV)** | **Area Ratio Hot/All** |
| 2 / $H_\beta$ | 0.23 | 0.139 | 1.840 | 21.9 | **0.197** |
| 2 / $H_\beta$ | 0.17 | 0.122 | 1.711 | 11.9 | **0.268** |
| 2 / $H_\beta$ | 0.15 | 0.130 | 2.033 | 26.0 | **0.224** |
| 2 / $H_\beta$ | 0.13 | 0.123 | 2.011 | 28.7 | **0.193** |
| 2 / $H_\beta$ | 0.11 | 0.131 | 2.020 | 34.5 | **0.241** |
| 2 / $H_\beta$ | 0.10 | 0.129 | 1.887 | 36.9 | **0.273** |
| 2 / $H_\beta$ | 0.08 | 0.127 | 1.643 | 43.2 | **0.331** |
| 2 / $H_\beta$ | 0.07 | 0.126 | 1.580 | 50.6 | **0.338** |
| 2 / $H_\beta$ | 0.06 | 0.127 | 1.643 | 54.5 | **0.258** |
| 2 / $H_\beta$ | 0.06 | 0.126 | 1.579 | 41.9 | **0.199** |
| 2 / $H_\beta$ | 0.05 | 0.110 | 1.780 | 31.4 | **0.084** |
| 3 / $H_\beta$ | 0.22 | 0.136 | 1.359 | 21.0 | **0.115** |
| 3 / $H_\beta$ | 0.18 | 0.116 | 1.325 | 13.6 | **0.123** |
| 3 / $H_\beta$ | 0.15 | 0.140 | 1.515 | 18.6 | **0.069** |
| 3 / $H_\beta$ | 0.13 | 0.129 | 1.639 | 27.9 | **0.063** |
| 3 / $H_\beta$ | 0.11 | 0.131 | 1.647 | 33.2 | **0.106** |
| 3 / $H_\beta$ | 0.09 | 0.121 | 1.747 | 27.1 | **0.120** |
| 3 / $H_\beta$ | 0.08 | 0.113 | 1.898 | 39.8 | **0.181** |
| 3 / $H_\beta$ | 0.07 | 0.118 | 1.881 | 43.3 | **0.284** |
| 3 / $H_\beta$ | 0.05 | 0.120 | 1.746 | 53.0 | **0.338** |
| 3 / $H_\beta$ | 0.04 | 0.120 | 1.850 | 49.9 | **0.304** |
| 3 / $H_\beta$ | 0.03 | 0.122 | 1.836 | 45.6 | **0.119** |

Table 6. Analysis of Hβ lines in 200 W water plasmas

| RF Power = 200W | | | | | |
|---|---|---|---|---|---|
| Position / Balmer Lines | Pressure (torr) | Doppler Energy Cold H (eV) | Doppler Energy Warm H (eV) | Doppler Energy Hot H (eV) | Area Ratio Hot/All |
| 2 / H$_\beta$ | 0.22 | 0.131 | 1.734 | 13.6 | **0.228** |
| 2 / H$_\beta$ | 0.18 | 0.127 | 1.920 | 21.1 | **0.223** |
| 2 / H$_\beta$ | 0.16 | 0.129 | 1.930 | 23.9 | **0.247** |
| 2 / H$_\beta$ | 0.14 | 0.125 | 2.046 | 33.4 | **0.226** |
| 2 / H$_\beta$ | 0.12 | 0.100 | 1.901 | 35.2 | **0.257** |
| 2 / H$_\beta$ | 0.10 | 0.123 | 1.913 | 29.7 | **0.219** |
| 2 / H$_\beta$ | 0.08 | 0.126 | 1.870 | 42.2 | **0.311** |
| 2 / H$_\beta$ | 0.07 | 0.129 | 1.837 | 52.7 | **0.341** |
| 2 / H$_\beta$ | 0.05 | 0.139 | 1.471 | 46.7 | **0.380** |
| 2 / H$_\beta$ | 0.04 | 0.125 | 1.594 | 58.9 | **0.308** |
| 2 / H$_\beta$ | 0.03 | 0.113 | 1.740 | 54.5 | **0.164** |
| 3 / H$_\beta$ | 0.15 | 0.124 | 1.550 | 25.1 | **0.152** |
| 3 / H$_\beta$ | 0.13 | 0.120 | 1.719 | 39.8 | **0.174** |
| 3 / H$_\beta$ | 0.11 | 0.122 | 1.812 | 52.0 | **0.208** |
| 3 / H$_\beta$ | 0.09 | 0.122 | 1.917 | 62.7 | **0.383** |
| 3 / H$_\beta$ | 0.07 | 0.121 | 1.750 | 61.4 | **0.479** |
| 3 / H$_\beta$ | 0.05 | 0.120 | 1.793 | 54.5 | **0.425** |
| 3 / H$_\beta$ | 0.04 | 0.122 | 1.877 | 56.3 | **0.400** |
| 3 / H$_\beta$ | 0.03 | 0.119 | 1.700 | 42.0 | **0.251** |
| 3 / H$_\beta$ | 0.02 | 0.127 | 1.549 | 15.3 | **0.156** |

Figure Captions

FIGURE 1. Hydrogen Balmer α lines indicate presence of hydrogen atoms with energy in excess of 40 eV. (a) At 0.08 Torr, 150 W, Position 3, Doppler broadening from H atoms with energies greater than 40 eV is seen at the base of the Hα peak. No 'hot' hydrogen is seen at higher pressures. (b) Fitting the lines requires three peaks, for 'cold' (<1 eV), 'warm' (<2.5 eV) and 'hot' hydrogen. (c) Between the electrodes in control plasmas (150 W, 0.05 Torr, $H_2$/Xe,20:1) only cold and warm hydrogen are found. Away from the electrode region only cold hydrogen is present.

FIGURE 2. Hα broadening indicates 'hot' atomic hydrogen energy a function of pressure but not position. (a) At 100 W the broadening is approximately 40 eV up to a pressure of 0.08 Torr. (b) At 150 W the measured broadening is approximately 45 eV up to a pressure of about 0.08 Torr. (c) At 200 W the measured broadening is above 45 eV up to a pressure of nearly 0.10 Torr. In all cases the 'warm' hydrogen is less than 2eV and the 'cold' hydrogen line width is so small it probably reflects factors (natural line width, Stark effect, instrument effects, etc.) other than Doppler broadening. Also note that the 'hot' hydrogen Doppler energy drops to between 10 and 20 eV at pressures above 0.10 Torr.

FIGURE 3. Hβ broadening also indicates 'hot' atomic hydrogen energy a function of pressure, but not position. (a) At 100 W the broadening is greater than 40 eV up to a pressure of 0.06 Torr. (b) At 150 W the measured broadening is greater than 40 eV up to a pressure of about 0.09 Torr. (c) At 200 W the measured broadening is above 40 eV up to a pressure of nearly 0.10 Torr. In all cases the 'warm' hydrogen is less than 2eV and the 'cold' hydrogen line width is so small it probably reflects factors (natural line width, Stark effect, instrument effects, etc.) other than Doppler broadening. Also note that the 'hot' hydrogen Doppler energy drops to between 10 and 20 eV at pressures above 0.10 Torr.

FIGURE 4. 'Hot' hydrogen between 10% and 45% of all atomic hydrogen. (a) and (b) At a position 15 cm from the electrode the fraction hot hydrogen varies as a function of pressure, and power to a lesser extent. Virtually identical trends are observed from the Hα and Hβ data. (c) and (d) The trends in fraction hot hydrogen as a function of pressure and power are virtually identical between the plates and at 15 cm from the plates.

FIGURE 5. (a) Excitation energy (Tex) determined as a function of power, position and hydrogen species used to determine peak intensity. (b) Boltzman plot showing the high fidelity of the data. Clearly computing excitation temperature on the 'cold' or 'hot' component of the line intensity makes only a small difference.

FIG. 1

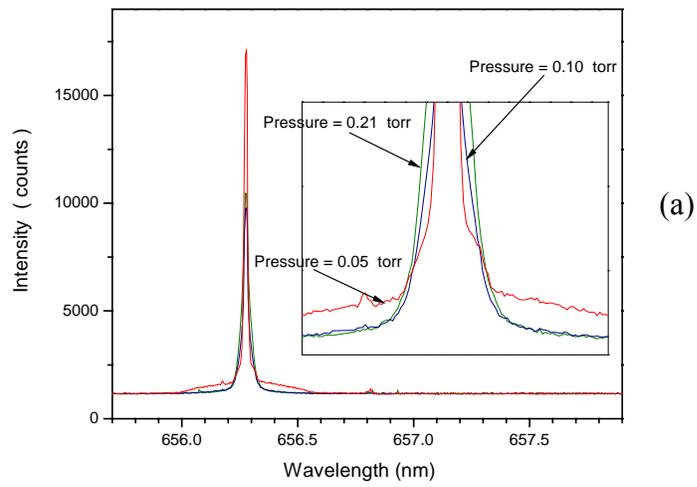

(a)

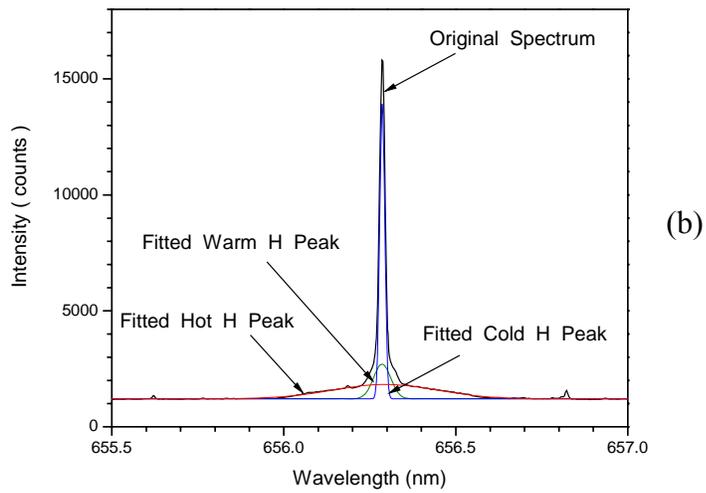

(b)

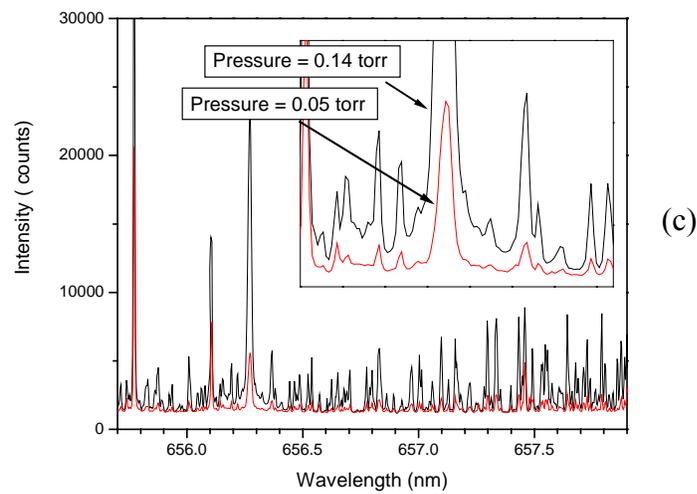

(c)

FIG. 2

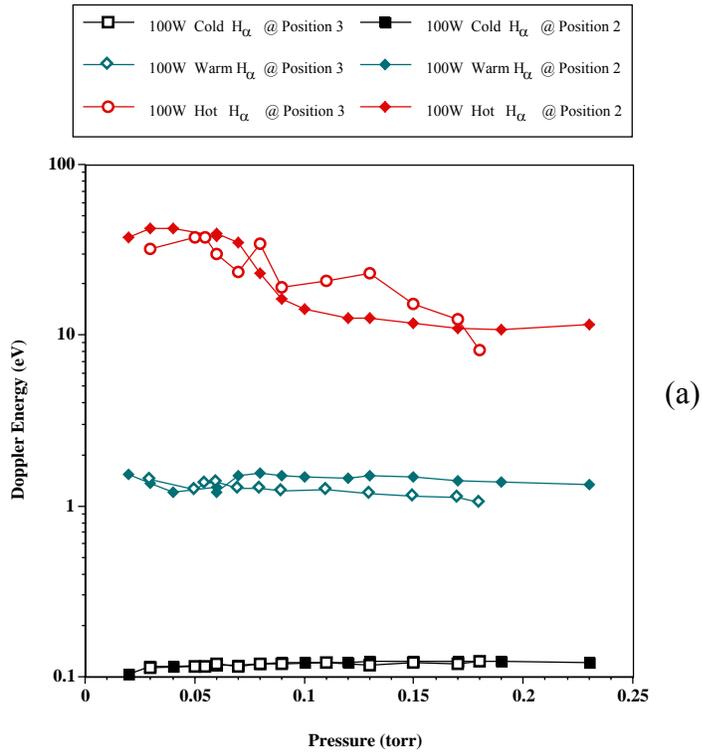

(a)

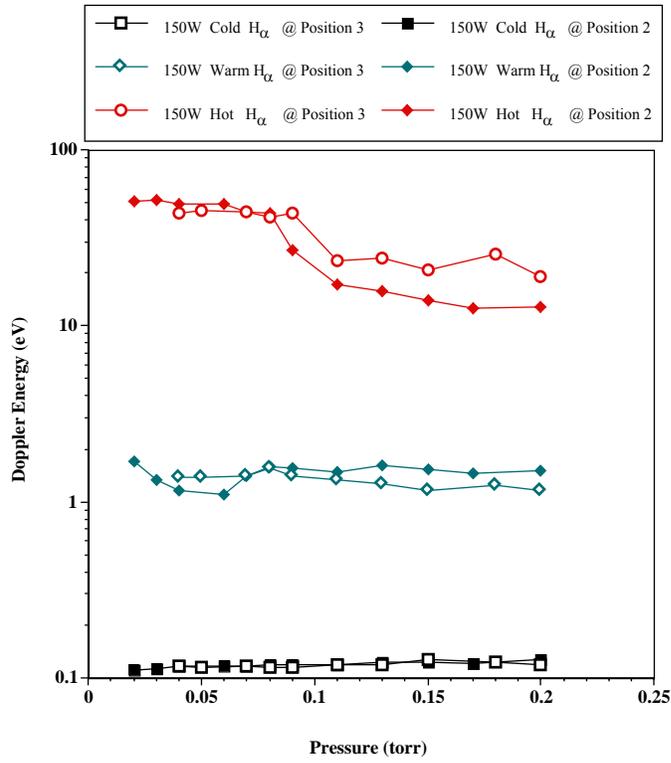

(b)

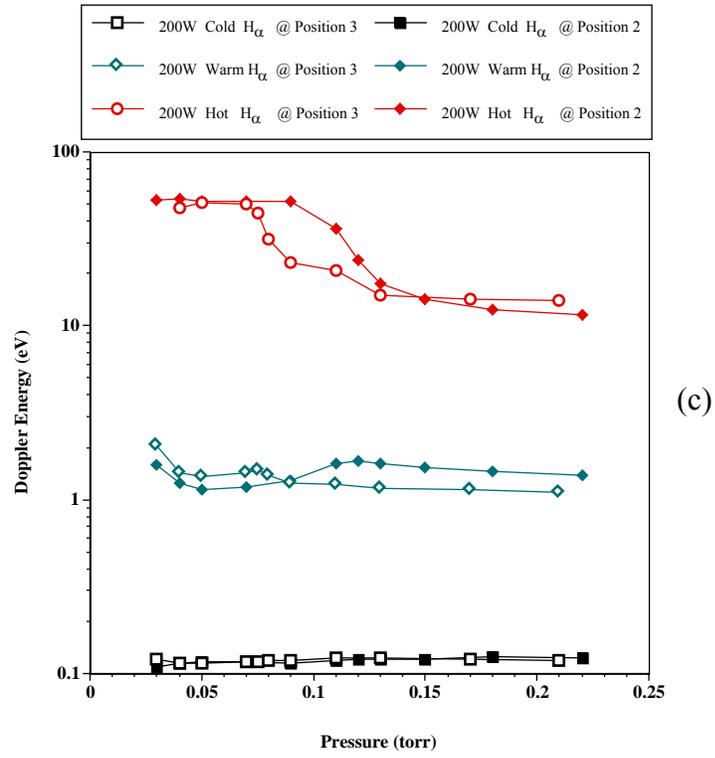

FIG. 3

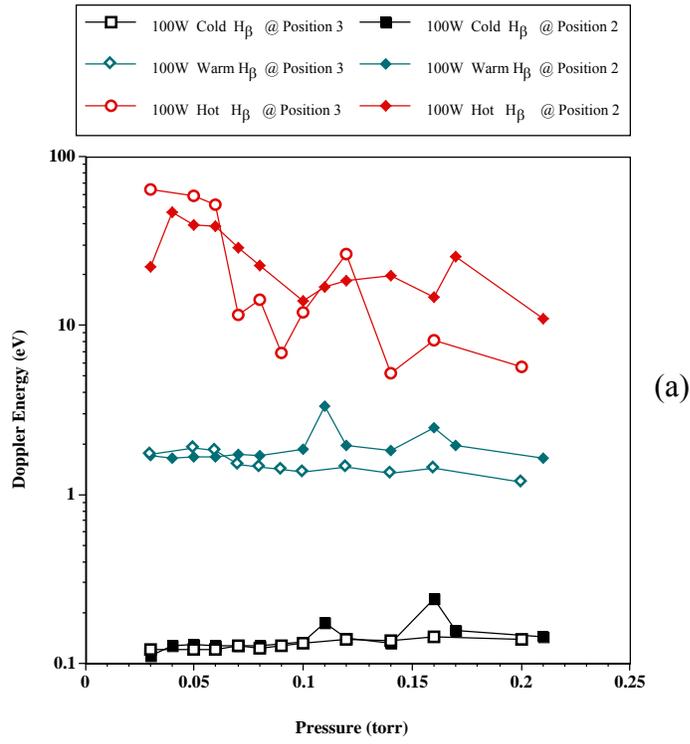

(a)

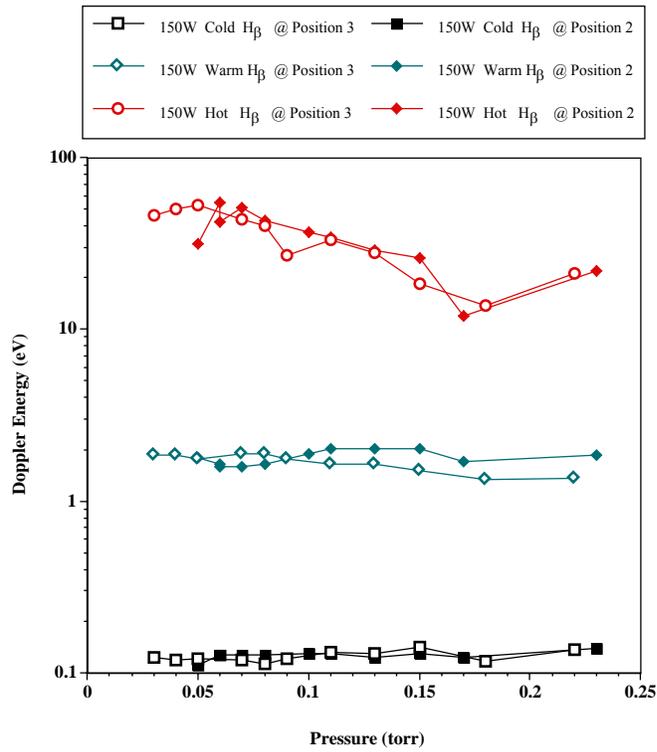

(b)

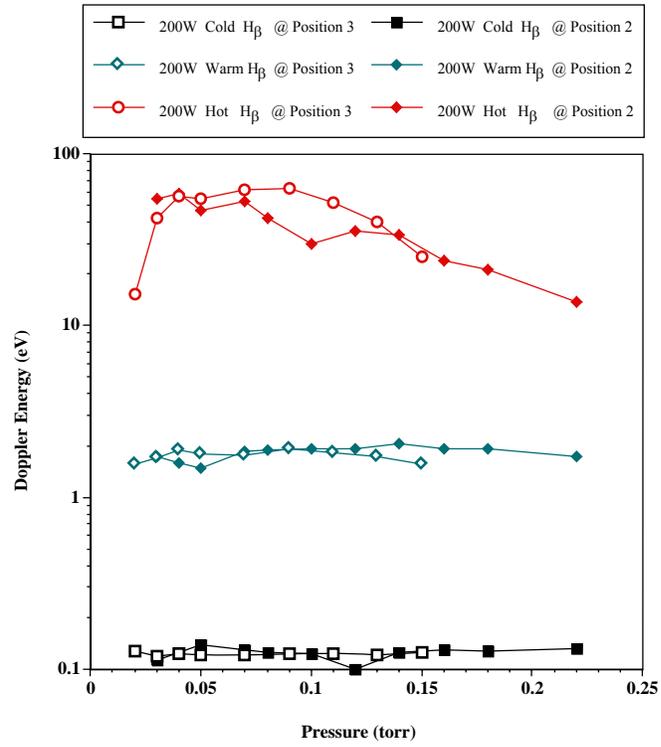

FIG. 4

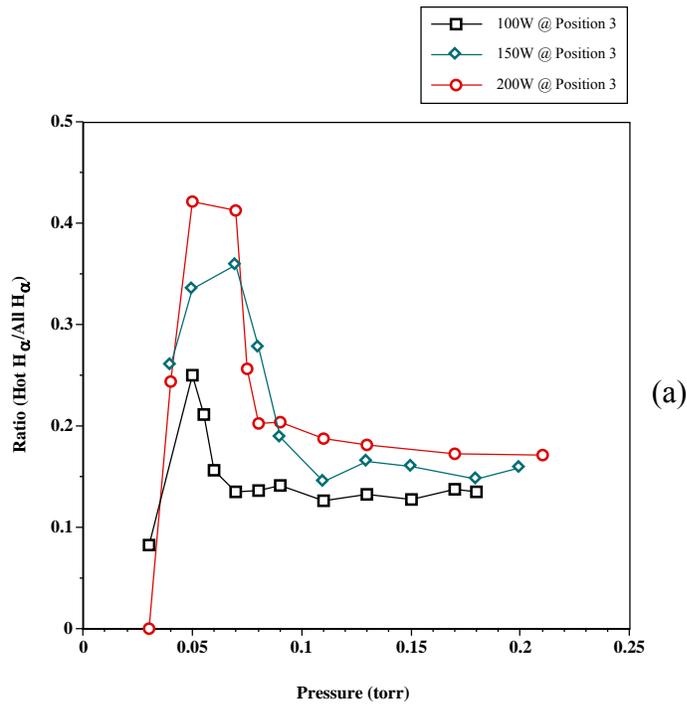

(a)

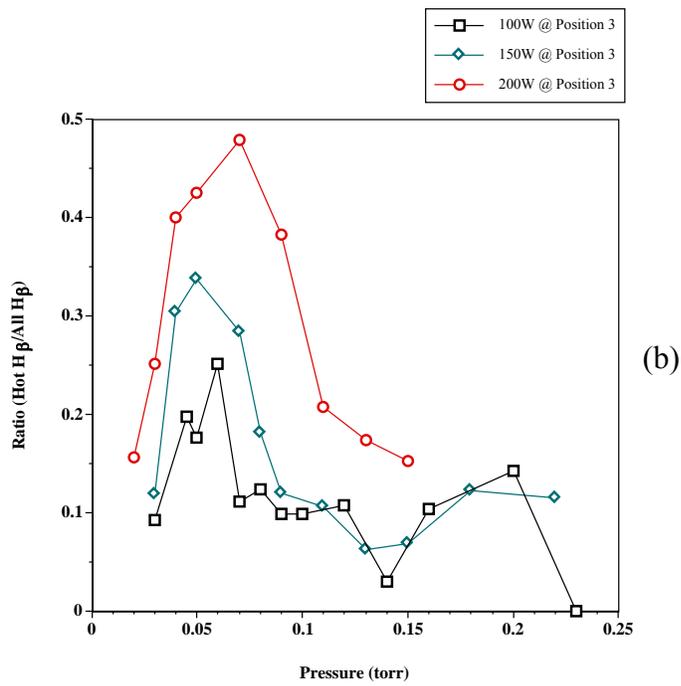

(b)

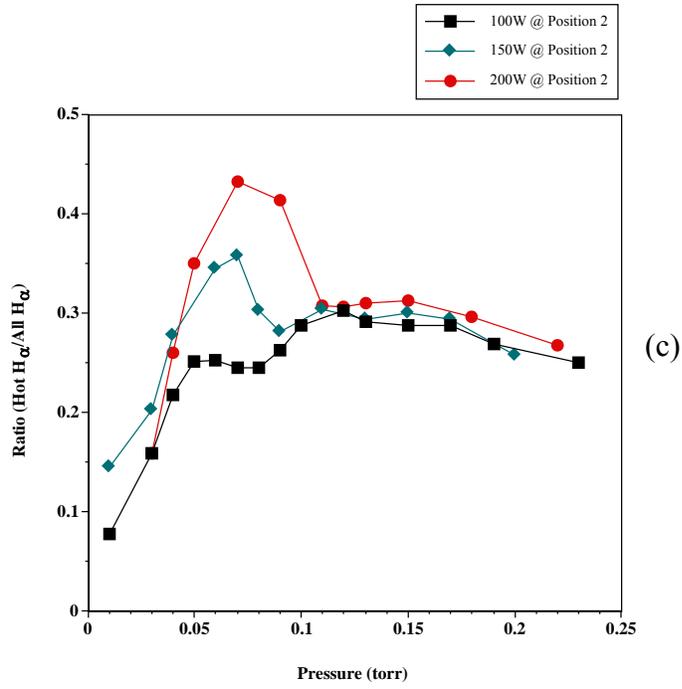

(c)

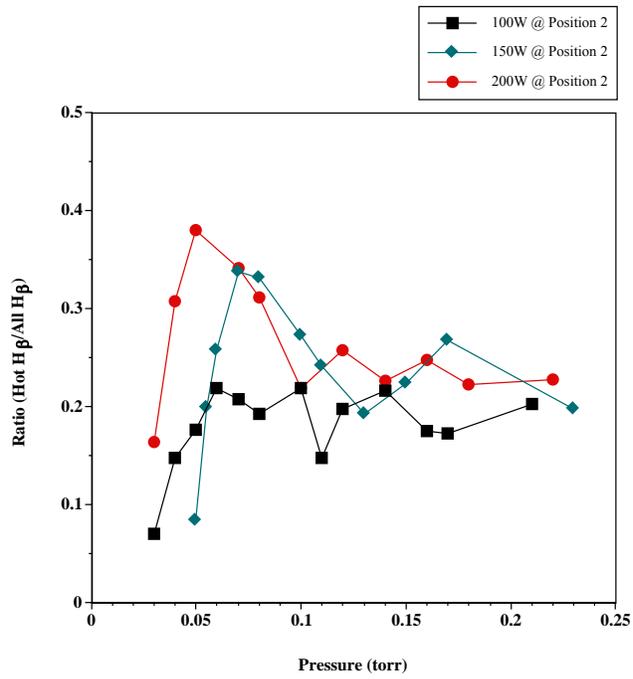

(d)

FIG. 5

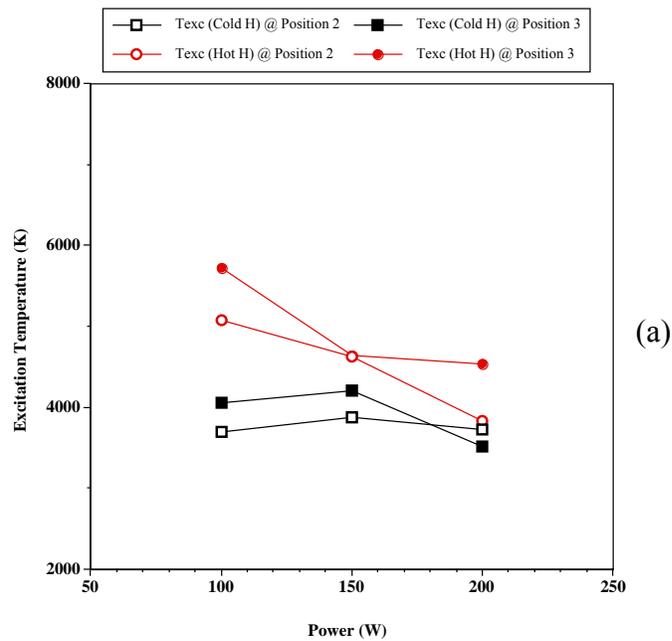

(a)

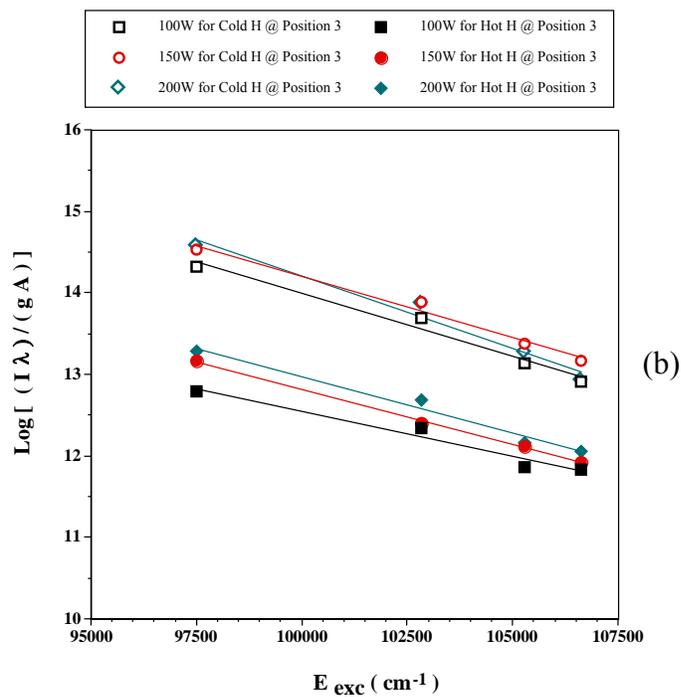

(b)